\documentstyle[11pt,epsf]{article}
\def\refer#1#2#3#4#5{#1, #2 {\bf #3}, #5 (#4).}
\newcommand{\nl}{\nonumber \\}
\newcommand{\bea}{\begin{eqnarray}}
\newcommand{\eea}{\end{eqnarray}}
\def\ra{\rightarrow}
\def\mab{{\cal M}_{ab}}
\def\mr{{\cal M}_R}
\def\md{{\cal M}_D}
\def\sigab{E\frac{d\sigma_{ab}}{dM^2d^3P}}
\def\sigr{E\frac{d\sigma_{R}(M)}{d^3P}}
\def\dphiab{\frac{d^3p_a}{2E_a}\frac{d^3p_b}{2E_b}}
\def\gabm{\Gamma_{R\ra a+b}(M)}
\def\spvmu{\left(g^{\mu\mu^\prime}-\frac{P^\mu P^{\mu^\prime}}{M^2}\right)}
\def\spvnu{\left(g^{\nu\nu^\prime}-\frac{P^\nu P^{\nu^\prime}}{M^2}\right)}
\def\gamab{E\frac{d\Gamma_{1\ra a+b+X}}{dM^2d^3P}}
\def\gamr{E\frac{d\Gamma_{1\ra R+X}(M)}{d^3P}}
\def\be{\begin{equation}}
\def\ee{\end{equation}}
\begin{document}
\title{{
Are the production and decay of a resonance always 
independent?}}
\author{Peter Lichard\footnote{
Dedicated to J\'{a}n Pi\v{s}\'{u}t on the occasion 
of his 60th birthday.} \footnote{On leave of absence from Department of 
Theoretical Physics, Comenius University,
842-15 Bratislava, Slovak Republic.}\\
Institute of Physics, Silesian University, 746-01 Opava, Czech Republic\\
and\\
Department of Physics and Astronomy, University of Pittsburgh, \\ Pittsburgh,
Pennsylvania 15260, USA}
\maketitle
\begin{abstract}
The widely accepted 
assumption that the decay of a resonance proceeds
independently of its production, quantitatively expressed as a factorizing 
formula for the differential cross section in the invariant mass of
unpolarized resonance debris, is put under scrutiny. It is shown that 
the factorization is always valid for scalar and pseudoscalar resonances,
although the usual version of the  formula is not entirely correct.
For resonances with nonzero spins the factorization does not generally take 
place. We deal in more detail with the spin-one case, where we show 
a condition on the decay matrix element that ensures the validity of the 
same factorizing formula as in the spinless case. This condition is 
satisfied for $\rho\ra\pi\pi$ but not, e.g., for $K^*\ra\pi K$ or 
$a_1\ra\pi\rho$. The formalism is applied also to the case when the
resonance is produced not in two-body collisions but in the decay 
of a heavier particle or resonance (chain decay). Application of
our formulas to the $e^+e^-$ production in one-photon approximation
agrees with what is known from quantum electrodynamics and thus provides 
another test of their soundness.
\end{abstract}
\section{Introduction}
\label{intro}

It has been conjectured from the early sixties (see, e.g., 
\cite{jackson64,pisutroos}) that a process in which some of
the final state particles, let us say $a$ and $b$, appear as decay 
products of a resonance $R$ can be viewed as a sequence of two independent 
steps:
(i) a reaction in which the resonance (treated as a stable particle
with mass $M$) together with other final-state particles is produced; 
(ii) the decay of the resonance, $R\ra a+b$.\footnote{We will consider
only two-body decay modes to keep the notation transparent. Our conclusions
are valid for any decay mode.}
This notion has been expressed quantitatively by factorizing the cross 
section for producing the final system that includes resonance decay
products (integrated over the angles in their common rest frame) with
the invariant mass from interval $(M,M+dM)$ as the product of
the cross section for producing the resonance with a (generally 
``off-shell") mass $M$ and a function describing its decay 
\cite{pilkuhn}
\be
\label{factorize}
d\sigma_{ab}=d\sigma_R(M)\left[\frac{\pi^{-1}M_0\gabm}
{\left(M_0^2-M^2\right)^2+M_0^2\Gamma^2(M)}\right]\ dM^2\ .
\ee
Here, $M_0$ is the nominal mass of the resonance, $\gabm$ is the
partial decay rate, and $\Gamma(M)$ is the mass dependent total decay width. 

We will argue in what follows that this formula is a little incorrect.
The correct formula, which will be derived in Sec.~\ref{derivation},
contains $M$ instead of $M_0$ in the numerator.

For the one-decay-mode resonances, the numerator and denominator in 
(\ref{factorize})
contain the same $\Gamma(M)$. The latter quantity was treated differently 
by different authors. Jackson \cite{jackson64}
related it to the decay rate of the resonance, which can be
expressed in terms of the $S$-matrix element and evaluated if
the interaction Lagrangian among the resonance and its decay products
is postulated. On the contrary, Pi\v{s}\'{u}t and Roos \cite{pisutroos},
who dealt with the $\rho(770)$ resonance, considered $\Gamma(M)$ 
a phenomenological object which can and should be determined by 
fitting the experimental data.
 
The idea of the cross section factorization arose in the isobaric nucleon
model \cite{lindster57,bergia60}. Although Eq. (\ref{factorize}) has 
never been proven generally, it has widely been accepted and heavily 
used because it matches the natural physical expectation. As far as we 
know, it was first introduced in its one-mode version  in
\cite{jouvet63}, as a generalization based on the cross sections for 
several reactions evaluated in the lowest order of the perturbation 
expansion using simple Lagrangians \cite{bordes63,bordjou63,abillon63}.

An attempt to prove Eq. (\ref{factorize}) was done in the textbook
\cite{pilkuhn}. Unfortunately, as we show below, in a not fully
correct way and, because of the choice of the propagator, only for 
spinless resonances.

In this paper we show that the factorization of the cross section
in the sense of Eq. (\ref{factorize}) is not as trivial as the naive
physics intuition suggests and that it is not valid generally.
We prove that for the spinless resonances the cross section really 
factorizes, although the correct formula differs a little from 
(\ref{factorize}). For spin-one resonances we get the factorizing
formula only if an additional assumption is made about the matrix element
for the $R\ra a+b$ decay. This assumption represents a severe constraint
that excludes many well-known resonances ($K^*$ and $a_1$, for 
example).

The motivation for this study was twofold, experimental and theoretical. 
On the experimental side,
there is a longstanding unresolved discrepancy between the parameters
of the $a_1$ resonance determined from the $\tau$-lepton decay on
one side and from hadronic production processes on the other, see
\cite{pdg}, p. 380. It is true that the experimentally observed
invariant mass spectrum of the decay products may be influenced by the
resonance mass dependence of the production cross section already
in the case when the factorization takes place. But the discrepancy
among the observed widths of $a_1$ is too big, which casts doubts on
the validity of the general assumption that the decay of a resonance
is independent of the way the resonance has been produced. 

Also the mass spectrum of the $K\pi$ system resulting from the $K^*$
decay was found to be process dependent (see \cite{bellog94} and 
literature cited there). But if the factorization in the sense of
(\ref{factorize}) is correct and the smooth factor $d\sigma_R(M)$
is properly parametrized this dependence should not affect  the
values of resonance parameters $M_0$ and $\Gamma(M_0)$ determined
by fitting to the experimental data. It is the very essence of the 
factorization hypothesis that the resonance part of the cross
section formula is process independent, and all the properties
of the resonance production mechanism are hidden in $d\sigma_R(M)$. 

On the theoretical side, many papers have appeared that calculated the
cross sections and decay rates of the processes in which a meson 
resonance emerges as one of the final state particles. What follows
are a few examples rather than the complete list:
$\phi\ra\rho\pi$ \cite{gomm84,songko95}, 
$D(1285),\  E(1420)\ra \rho\pi\pi$,
$D\ra a_1 \pi$, $Q\ra K^*\pi\pi$
\cite{gomm84},
$\eta^\prime\ra\rho\gamma$
\cite{durso87},
$\pi\pi\ra\rho\gamma$
\cite{jkplds91},
$a_1\ra\rho\pi$
\cite{xiong92,song93,haglin94},
$f_2\ra \rho\gamma,\  \rho\pi\pi$
\cite{suzuki93},
$K_1\ra K\rho$, $b_1\ra\pi\omega$
\cite{haglin94}.
In these calculations, the outgoing resonances were treated as
stable particles with sharp masses. It is not clear to what extent 
such an approximation is justified. It is obvious that it
cannot give reliable results in cases when the available 
phase space is small due to a close proximity to the kinematic threshold.
A typical example is $K_1\ra K\rho$, where the available
energy is less than 10 MeV, much smaller than the widths of
participating resonances. There is also at least one case in which the 
sharp resonance mass approach is not applicable at all, namely the 
$K_1(1270)\ra K\omega$ decay. This decay has a relatively big 
branching ratio $(11.0\pm2.0)\%$ \cite{pdg} although the nominal mass of
the $K_1$ resonance is less than the sum of the $K$ and $\omega$ masses. 

In simple processes that do not suffer from severe phase space 
limitation there is no obvious reason for disregarding the sharp
resonance mass method. The question of how well the method works under 
those circumstances naturally arises. Answering it by comparing
with experimental data is usually impossible not only because
of nonexistent or statistically limited data but also because  
additional theoretical assumptions are usually made simultaneously. 
Thus only one way remains--to compare the sharp resonance mass method 
with a more complete calculation that involves also the last step--the 
conversion of the produced resonance to the hadrons that are really observed
experimentally. Of course, the calculation involving this step is more
difficult. But if the factorization takes place
there is no need to go through the complete procedure
for each process considered. We only need to know  the decay rate 
or cross section for producing the resonance with arbitrary mass in the sharp
mass approximation. The conversion of the resonance to its decay products
is then described by a simple universal (independent of the dynamics 
of the production process) function. This simplifies the evaluation
of some of the processes mentioned above greatly.

The paper is organized as follows: In Sec. \ref{derivation} we
present the derivation of the factorizing relation for the cross
section in the spinless case  and show a sufficient condition
under which it is valid also for unit spin resonances.
Processes in which a resonance itself appears as a decay product 
are dealt with in Sec.~\ref{decays}.
Sec.~\ref{qed} is devoted to the connection of our formalism with methods 
used in quantum electrodynamics.
In Sec.~\ref{spinone} we investigate how is the factorization condition 
for spin-one resonances fulfilled in some important cases.
Our main conclusions are summarized and commented upon in Sec.~\ref{comments}.

\section{Derivation of the cross section formula}
\label{derivation}
\subsection{Generalities}
\label{generali}

Let us consider a two-body reaction (see Fig.~1)
\be
\label{reactab}
1+2\ra a+b+X
\ee
with particles  $a$ and $b$ originating from the decay of a resonance $R$
and $X$ representing the system of $n$ other outgoing particles. We will 
assume that none of the latter is identical with either $a$ or $b$. 
We will be interested in the cross section for producing the $(a,b)$ 
system with fixed invariant mass $M$, $M^2=(p_a+p_b)^2$, and fixed 
three-momentum ${\bf P}={\bf p}_a+{\bf p}_b$. For $a$ and $b$ the sum 
over the spin states is assumed together with integration over the 
momenta in their common rest frame.

Concerning the particles from the $X$-system, we will write our 
formulas for the cross section integrated over their momenta and summed
over their spin states. But our conclusions would remain unchanged if 
we opted for a more detailed differential cross section. For simplicity,
we will not display the dependence of  matrix elements on momenta and
polarizations of particles from the $X$-system and will use the simplified
notation $\lambda_X$ for the complete set of their polarization indices.
For reader's convenience we start with the general cross section formula
\be
\label{cross}
d\sigma=\frac{(2\pi)^4}{4F}\left|{\cal M}\right|^2\delta^{(4)}
(p_1+p_2-\sum_i p_i)\prod_i\frac{d^3p_i}{(2\pi)^32E_i}\ ,
\ee
where the sum and product run over the final state particles and
$F=E_1E_2\left|{\bf v}_1-{\bf v}_2\right|$.\footnote{This formula is more
general than that given in \cite{pdg}. The latter applies only to
head-on collisions \cite{ll2}, in which $F=p_{1cm}\sqrt{s}=m_2p_{1lab}$.} 
Using (\ref{cross}) and 
\be
\label{delta2}
2E\ \delta(M^2-(p_a+p_b)^2)\ \delta({\bf P}-{\bf p}_a-{\bf p}_b)=
\delta^{(4)}(P-p_a-p_b)
\ee
it is easy to show that the cross section of reaction (\ref{reactab})
we are dealing with is given by
\bea
\label{cross_ab}
\sigab&=&\frac{1}{32\pi^2F}
\int\sum_{\lambda_X,\lambda_a,\lambda_b}\left|{\cal M}_{ab}\right|^2
\delta^{(4)}(p_1+p_2-P_X-P)\\
&\times& \delta^{(4)}(P-p_a-p_b)\ \dphiab\ d\Phi_X   \ ,\nonumber
\eea
where
\be
\label{dPhiX}
d\Phi_X=\prod_{i\in X}\frac{d^3p_i}{(2\pi)^3 2E_i}\ .
\ee
For further considerations it is necessary to introduce the sharp mass
approximation. We define a particle $R(M)$ with the quantum numbers identical
to those of the resonance $R$, but with a fixed mass $M$, which may be 
different from the nominal resonance mass $M_0$. The cross section of the
reaction
\be
\label{reactr}
1+2\ra R(M) + X\ ,
\ee
in which the unpolarized $R(M)$ with momentum $P$ is produced 
(see Fig.~2), is given by 
\be
\label{cross_R}
\sigr=\frac{\pi}{4F}
\int\sum_{\lambda_X,\lambda_R}\left|{\cal M}_R\right|^2
\delta^{(4)}(p_1+p_2-P_X-P)d\Phi_X
\ee
Now, let us turn to the decay 
\be
\label{decay}
R(M) \ra a+b \ ,
\ee
depicted in Fig.~3.
The general formula for the differential decay rate of a particle
with mass $M$ into $n$ daughter particles reads \cite{pdg}
\be
\label{dgamma}
d\Gamma=\frac{(2\pi)^4}{2M}\left|{\cal M}\right|^2\delta^{(4)}
(P-\sum_{i=1}^n p_i)\prod_{i=1}^n\frac{d^3p_i}{(2\pi)^3 2E_i}\ .
\ee
On the basis of it we can write the decay rate of (\ref{decay})
averaged over the spin states of $R(M)$ and summed over the final state
ones as
\bea
\label{decayrate}
\Gamma_{R\rightarrow a+b}(M)&=&\frac{1}{8(2J_R+1)M\pi^2}
\int\sum_{\lambda_R,\lambda_a,\lambda_b}\left|{\cal M}_D\right|^2\\
&\times&\delta^{(4)}(P-p_a-p_b)
\dphiab \ . \nonumber
\eea
Here, $J_R$ denotes the spin of the resonance.

\subsection{Formula for spinless resonances}
\label{spinless}
For a scalar or pseudoscalar resonance the matrix element for
reaction (\ref{reactab}) is given by
\be
\label{mabscalar}
\mab=\mr {\cal P}(P) \md\ ,
\ee
where the propagator of the spin-zero resonance is written in the
form\footnote{We will change notation of $\Gamma(P^2)$ to $\Gamma(M)$
later on.} 
\be
\label{scalarprop}
{\cal P}(P)=\frac{i}{P^2-M_0^2+iM_0\Gamma(P^2)}
\ee
to which the bare propagator of spinless particle develops after 
convoluting with one-particle-irreducible bubbles and summing over all
such diagrams.
Substitution of (\ref{mabscalar}) into (\ref{cross_ab}) gives
\bea
\sigab&=&\frac{\left|{\cal P}(P)\right|^2}{32\pi^2F}
\int\sum_{\lambda_X}\left|\mr\right|^2
\delta^{(4)}(p_1+p_2-P_X-P)d\Phi_X \nl
&\times& \int\sum_{\lambda_a,\lambda_b}\left|\md\right|^2
\delta^{(4)}(P-p_a-p_b)\dphiab \nonumber
\eea
The integrals can be expressed in terms of observable quantities
using (\ref{cross_R}) and (\ref{decayrate}) with $J_R=0$ and no 
summing over $\lambda_R$. The result is
\be
\label{myformula}
\sigab=\sigr
\left[\frac{\pi^{-1}M\Gamma_{R\ra a+b}(M)}
{\left(M_0^2-M^2\right)^2+M_0^2\Gamma^2(M)}\right]\  .
\ee 
Our formula (\ref{myformula}) differs from the formula (\ref{factorize}) 
used in analysing the resonance production experiments up to now.
The fixed $M_0$ in numerator is replaced by variable $M$. To localize the
source of discrepancy we went through the derivation in textbook 
\cite{pilkuhn}. We suspect that the factor $m_d$ in the right-hand side
of Eq. (8.7) on p. 166 should be replaced by $s_d^{1/2}$ (there is no 
other $m_d$, either explicit or hidden, in that equation). After this
modification, Eq. (\ref{factorize}) agrees with (\ref{myformula}).

\subsection{Case of spin-one resonances}
\label{caseone}

The matrix element of the decay (\ref{decay}) can be written as
\be
\label{me1}
{\cal M}_D(P,\lambda_R)=B_\nu(P)\epsilon^\nu(P,\lambda_R)\ ,
\ee
where $P$ is the four-momentum of the resonance $R(M)$ and 
$\epsilon(P,\lambda_R)$ is its polarization vector. The four-vector $B$
contains all the dynamics of the $Rab$ interaction. We suppressed the momenta 
and possible polarization indices  of particles $a$ and $b$.
The sum of the matrix elements squared over possible spin states of
the resonance, which enters the formula for the decay rate of 
unpolarized $R(M)$'s, yields
\be
\label{me2}
\sum_{\lambda_R}\left|{\cal M}_D(P,\lambda_R)\right|^2=-
\spvnu B_\nu(P)B_{\nu^\prime}^*(P)\ .
\ee
With its help we can cast the decay rate formula (\ref{decayrate}) into
\be
\label{decayrate1}
\gabm=\frac{1}{24M\pi^2}\spvnu T_{\nu\nu^\prime}(P)\ ,
\ee
where we have defined the tensor
\be
\label{tmunu}
T_{\nu\nu^\prime}(P)=\int
\sum_{\lambda_a,\lambda_b}{\cal B}_{\nu\nu^\prime}(P)
\ \delta^{(4)}(P-p_a-p_b)\ \dphiab\ 
\ee
with
\be
\label{bnunu}
{\cal B}_{\nu\nu^\prime}=-\sum_{\lambda_a,\lambda_b}B_\nu
B_{\nu^\prime}^*\ .
\ee
The tensor (\ref{tmunu}) cannot depend  on anything else but four-vector $P$ 
because $p_a$ and $p_b$ are integrated out and the sum over spin states
is performed.
   
Similarly, writing the matrix element for the reaction (\ref{reactr})
in the form
\be
\label{me1r}
{\cal M}_R(P,\lambda_R)=A_\mu(P)\epsilon^\mu(P,\lambda_R)\ ,
\ee
we arrive at the following cross section formula 
\bea
\label{crossr1}
\sigr&=&\frac{\pi}{4F}\spvmu \\
&\times&\int \sum_{\lambda_X}\left(-A_\mu 
A_{\mu^\prime}^*\right) \delta^{(4)}(p_1+p_2-P_X-P)\ d\Phi_X \nonumber
\eea
Now we are ready to attack the formula for the cross section of the
reaction (\ref{reactab}). The matrix element of the latter is given by
\be
\label{me1ab}
\mab=A_\mu{\cal P}^{\mu\nu}(P)B_\nu \ ,
\ee
where ${\cal P}^{\mu\nu}$ is the propagator of the spin-one resonance. 
Following \cite{isgur} we write it near the resonance mass ($P^2\approx
M^2_0$) as
\be
\label{vectorprop}
{\cal P}^{\mu\nu}(P)=-i\frac{g^{\mu\nu}-w(P^2){P^\mu P^\nu}/{P^2}}
{P^2-M_0^2+iM_0\Gamma(P^2)}\ .
\ee
The scalar function $w(P^2)$ reflects the properties of the 
one-particle-irreducible bubble. Fortunately, 
as we will see, under 
the assumption that  enables us to write formula (\ref{myformula}) also 
in the spin-one case, the term containing it will not contribute. 

Making use of (\ref{me1ab}) we may rewrite (\ref{cross_ab}) into the
form
\bea
\label{cross_ab1}
\sigab&=&\frac{1}{32\pi^2F}{\cal P}^{\mu\nu}(P){\cal P}^{*{\mu^\prime}
{\nu^\prime}}\ T_{\nu{\nu^\prime}}\\ 
&\times&
\int \sum_{\lambda_X}
\left(-A_\mu A_{\mu^\prime}^*\right)\delta^{(4)}(p_1+p_2-P_X-P)d\Phi_X
\eea
Without further assumptions, the right-hand side cannot be converted 
into a simple product of two factors, one describing the production of
a resonance, the other its decay. 

To proceed further let us assume that, as a result of a special dynamics 
of the $R\ra a+b$ transition, the tensor $T_{\nu\nu^\prime}(P)$, defined in 
(\ref{tmunu}), satisfies the relations  
\be
\label{gi}
P^\nu T_{\nu\nu^\prime}(P)= T_{\nu{\nu^\prime}}(P)P^{\nu^\prime}=0\ .
\ee
As an immediate consequence of our assumption we can simplify
(\ref{cross_ab1}) to
\bea
\label{cross_ab2}
\sigab&=&\frac{1}{32\pi^2F}\left|{\cal P}(P)\right|^2
T^{\mu{\mu^\prime}}\\ 
&\times& \int \sum_{\lambda_X}
\left(-A_\mu A_{\mu^\prime}^*\right)\delta^{(4)}(p_1+p_2-P_X-P)
\ d\Phi_X\ ,\nonumber
\eea
where ${\cal P}(P)$ is given by Eq. (\ref{scalarprop}). Condition (\ref{gi})
also signifies that the tensor $T^{\mu\mu^\prime}$ can be represented as
\be
\label{t2}
T^{\mu\mu^\prime}(P)=\spvmu T(M^2)\ ,
\ee
where $T$ is a scalar function of $M^2=P^2$. It can be determined
from
\be
\label{t3}
T(M^2)=\frac{1}{3}g^{\nu\nu^\prime}T_{\nu\nu^\prime}(P)\ .
\ee
The result is 
\be
\label{t4}
T(M^2)=8M\pi^2\gabm\ .
\ee
After inserting (\ref{t2}) with (\ref{t4}) into (\ref{cross_ab2}) and using
(\ref{crossr1}) we complete the proof that our central formula 
(\ref{myformula}) is valid also for spin-one resonances if condition
(\ref{gi}) is satisfied.

For practical calculations it is good to notice
that if the condition (\ref{gi}) is fulfilled then
the decay rate can be calculated from a simpler formula following
from (\ref{decayrate1}), namely
\be
\label{simplerate}
\Gamma_{R\rightarrow a+b}(M)=\frac{1}{24M\pi^2}
\int\ g^{\nu\nu^\prime}{\cal B}_{\nu\nu^\prime}\ \delta^{(4)}(P-p_a-p_b)
\dphiab \ .
\ee
Let us also note that it follows from (\ref{gi}) that 
\be
\label{gi2}
P^\nu P^{\nu^\prime}T_{\nu\nu^\prime}=0\ .
\ee
This formula thus represents the necessary condition for (\ref{gi}) 
be fulfilled.

\section{Resonances as decay products}
\label{decays}

In this section we deal with the case when a resonance $R$ appears itself
as one of the decay products of a heavier particle or resonance and then
converts to stable particles (see Fig.~4). We again call $a$ and 
$b$ the debris of the resonance $R$. On the basis of (\ref{dgamma}) and
(\ref{delta2}) we write the following formula for the differential decay 
rate 
\bea
\label{decay_ab}
\gamab&=&\frac{1}{16M\pi^2}
\int\sum_{\lambda_X,\lambda_a,\lambda_b}\left|{\cal M}_{ab}\right|^2
\delta^{(4)}(p_1-P_X-P)\\
&\times& \delta^{(4)}(P-p_a-p_b)\ \dphiab\ d\Phi_X   \ ,\nonumber
\eea
which is a decay analogue of Eq. (\ref{cross_ab}). The rate of the decay
in which the unpolarized sharp-mass resonance $R(M)$ is produced, 
see Fig.~5,
reads as
\be
\label{decay_R}
\gamr=\frac{\pi}{2M}
\int\sum_{\lambda_X,\lambda_R}\left|{\cal M}_R\right|^2
\delta^{(4)}(p_1-P_X-P)d\Phi_X
\ee
Repeating steps we have performed in Sec.~\ref{derivation} 
we can easily check that the following relation between the
decay rates of $1\ra a+b+X$ (Fig.~4) and $1\ra R(M)+X$
(Fig.~5) holds
\be
\label{mydecay}
\gamab=\gamr
\left[\frac{\pi^{-1}M\Gamma_{R\ra a+b}(M)}
{\left(M_0^2-M^2\right)^2+M_0^2\Gamma^2(M)}\right]\ .
\ee 
This relation is valid for all scalar and pseudoscalar resonances 
and for such decays of intermediate spin-one resonances that satisfy
condition (\ref{gi}). Again, spins of the parent particle $1$, of 
the particles from the $X$ system, and of particles $a$ and $b$  are
not important, but the sum over the $a$ and $b$  polarizations must 
be performed.

\section{Application to quantum electrodynamics}
\label{qed}

Up to now we have spoken about short-lived hadronic resonances in intermediate 
states of reactions and decays. But in the derivation of our key
formulas (\ref{myformula},\ref{mydecay}) what was important was the structure
of Feynman diagrams, not the type of interactions that cause the production
and decay to happen. Our results can be applied to all  strong,
electromagnetic and weak processes with the same pattern of Feynman
diagrams. 

In this section we are going to show that the so called invariant
integration method, which was often  used for the cross section
calculations in quantum electrodynamics \cite{invint}, can be regarded
as a special case of our formulas.

Let us consider a reaction in which a pair of oppositely charged leptons
is produced in one-photon approximation. See Fig.~1 with $R\ra
\gamma, a\ra e^+, b\ra e^-$. The reaction can be viewed as
a two step process: (i) the production of a massive photon with mass $M$,
which is different from its ``nominal" mass $M_0=0$; (ii) the decay of
that massive photon, which is described by Feynman diagram depicted in
Fig.~6. The decay matrix element is given by Eq.~(\ref{me1}) with
\be
\label{bqed}
B_\nu=e\overline{u}_{\lambda_b}(p_b)\gamma_\nu v_{\lambda_a}(p_a)\ .
\ee
Quantity (\ref{bnunu}) now becomes ($m$ is the electron mass)
\be
\label{bnunuqed}
{\cal B}_{\nu\nu^\prime}=
2e^2\left(m^2g_{\nu\nu^\prime} 
-2p_{a\nu}p_{b\nu^\prime}-2p_{b\nu}p_{a\nu^\prime}\right)\ 
\ee
and vanishes when contracted with $P^\nu$ or $P^{\nu^\prime}$. This 
guarantees that the condition (\ref{gi}) is fulfilled and we may use
Eqs.~(\ref{myformula}), (\ref{mydecay}), and (\ref{simplerate}).
The contraction of (\ref{bnunuqed}) with  metric tensor yields
the constant $4(M^2+2m^2)$. What remains to calculate in (\ref{simplerate})
is the well-known two-particle phase-space integral
\be
\label{phasespace}
\int\frac{d^3p_a}{2E_a}\frac{d^3p_b}{2E_b}\ \delta^{(4)}(P-p_a-p_b)
=\frac{\pi}{2}\sqrt{1-\frac{4m^2}{M^2}}\ .
\ee
Putting everything together we are getting the following decay rate of
the massive photon to an $e^+e^-$ pair
\be
\label{gammadecayrate}
\Gamma_{\gamma(M)\ra e^+e^-}=\frac{\alpha M}{12}
\left(1+\frac{2m^2}{M^2}\right)
\sqrt{1-\frac{4m^2}{M^2}}\ ,
\ee
where $\alpha=e^2/4\pi$ is the fine structure constant.
If we now explore our factorizing formula with the photon mass $M_0=0$
we are getting the well-known QED relation \cite{invint}
\be
\label{invintformula}
E\frac{d\sigma_{e^+e^-}}{dM^2d^3P}=
E\frac{d\sigma_{\gamma(M)}}{d^3P}
\frac{\alpha}{3\pi M^2}
\left(1+\frac{2m^2}{M^2}\right)
\sqrt{1-\frac{4m^2}{M^2}}\ .
\ee
Let us note that if we had used the usual factorizing formula
(\ref{factorize}) instead of our Eq. (\ref{myformula}) we would not
have obtained (\ref{invintformula}), but  zero.

\section{Spin-one resonances and their decays}
\label{spinone}

In this section we are going to investigate vector and axial-vector mesons 
and their main decay modes to see whether they satisfy the condition
that ensures the factorization of the cross section (\ref{myformula})
and decay rate (\ref{mydecay}).
\subsection{Vector resonance decaying to two pseudoscalar mesons}
\label{vpp}
The charge-invariant interaction among a vector field and two pseudoscalar 
fields is described by the Lagrangian density
\be
\label{vpplag}
{\cal L}_{VPP}=\frac{ig}{2}{\rm Tr}\left(V^\mu\phi_1^\dagger
\stackrel{\leftrightarrow}{\partial}_\mu\phi_2\right)\ + {\rm h.c.},
\ee
where $V$ is the matrix in isospin space of vector field operators 
and $\phi_1$ and $\phi_2$ are those of pseudoscalar fields. Evaluation
of the decay matrix element provides it in the form (\ref{me1}) with
\be
\label{vppb}
B_\nu=c\left(p_{a\nu}-p_{b\nu}\right)\ ,
\ee
where $p_a$ and $p_b$ are the four-momenta of outgoing mesons and
$c$ includes the coupling constant $g$ and possible constants arising 
from the isospin structure of (\ref{vpplag}). Contraction
of (\ref{vppb}) with the four-momentum of the parent vector meson gives
\be
\label{vpppb}
P^\nu B_\nu=c\left(m_a^2-m_b^2\right)\ .
\ee
For final-state mesons with equal masses we have $P^\nu B_\nu=0$, what
immediately implies that the factorization condition (\ref{gi}) is 
satisfied. For unequal masses we proceed further and get
\be
\label{h}
P^\nu P^{\nu^\prime}T_{\nu\nu^\prime}(P)=-c^2\pi\left|{\bf p}^*_a\right|
\left(m_a^2-m_b^2\right)^2\ .
\ee
The momentum in the decay rest frame was marked by asterisk. It is clear
from (\ref{h})
that the factorization condition (\ref{gi}) cannot be fulfilled for 
unequal masses of pseudoscalar mesons.
 
We have just shown that the factorization formula can be used for
the $\rho\ra\pi\pi$ decay, but not for decays of vector mesons to a pair
of unequally heavy pseudoscalar mesons, e.g., $K^*\ra K\pi$.

\subsection{Axial-vector resonance decaying to vector and pseudoscalar
mesons}
\label{avp}

The most general Lagrangian for interaction among the axial-vector (R),
vector (a), and pseudoscalar (b) mesons consists of three parts. For simplicity
and because some authors used simpler Lagrangians consisting of one of 
those terms only, we will study these parts individually. For simplicity,
we consider $A$, $V$, and $\phi$ as field operators of individual particles.
We also introduce a shorthand notation for the polarization vector
of the vector particle $\epsilon_a=\epsilon(p_a,\lambda_a)$.

In one of the first papers about the $a_1$ resonance in the $\tau$-lepton
decay \cite{pham} the following Lagrangian was used:  
\be
\label{phamlag}
{\cal L}=gA_\mu^\dagger V_\mu\ +{\rm h.c.}
\ee
Four-vector $B$, defined by (\ref{me1}), is now proportional to the 
polarization vector of the vector meson 
\be
B_\nu=g\epsilon_{a\nu}\ .
\ee
After a little algebra we obtain
\be
P^\nu P^{\nu^\prime}{\cal B}_{\nu\nu^\prime}=\frac{g^2}{4m_a^2}
\lambda(M^2,m_a^2,m_b^2) \ ,
\ee
where we have introduced the triangle function
$\lambda(x,y,z)=x^2+y^2+z^2-2xy-2xz-2yz$.  This result implies that
the expression $P^\nu P^{\nu^\prime}T_{\nu\nu^\prime}$ must acquire a
nonvanishing constant value, what leads to violation of (\ref{gi2}). 
Condition (\ref{gi}) cannot therefore be satisfied.

Another possibility of a simple Lagrangian, which corresponds to the vertex
factor used, e.g., in \cite{xiong92}, is
\be
\label{xionglag}
{\cal L}=gA^\dagger_\alpha\
\left(\frac{\partial V^\alpha}{\partial x_\beta}
-\frac{\partial V^\beta}{\partial x_\alpha}\right)
\frac{\partial\phi}{\partial x^\beta}
\ +{\rm h.c.}
\ee
It leads to
\be
B_\nu=g\left[\left(p_ap_b\right) \epsilon_{a\nu}-
\left(p_b\epsilon_{a\nu}\right)p_{a\nu}\right]\ .
\ee
Because the quantity 
\be
P^\nu P^{\nu^\prime}{\cal B}_{\nu\nu^\prime}=\frac{g^2}{4}
\lambda(M^2,m_a^2,m_b^2) 
\ee
comes out nonvanishing, we conclude, as in the previous case, that  
the factorization cannot take place.

Finally, from the Lagrangian
\be
\label{janslag}
{\cal L}=g
\frac{\partial A^\dagger_\alpha}{\partial x^\beta}
\left(\frac{\partial V^\alpha}{\partial x_\beta}
-\frac{\partial V^\beta}{\partial x_\alpha}\right)\phi\ +{\rm h.c.}
\ee
it follows that
\be
B_\nu=g\left[\left(p_aP\right) \epsilon_{a\nu}-
\left(P\epsilon_a\right)p_{a\nu}\right]\ .
\ee
Now we have $P^\nu B_\nu=0$ and the condition (\ref{gi}) for formula
(\ref{myformula}) being valid is satisfied.

Lagrangian (\ref{janslag}), which is the only Lagrangian leading
to factorizing cross section for the production of an axial-vector
resonance decaying to a vector meson and a pseudoscalar meson,
was utilized, for example, in meson exchange model for $\pi\rho$
scattering \cite{janssen}.

Unfortunately, there is strong theoretical \cite{gomm84,rudaz} 
and phenomenological \cite{gao} evidence that the correct 
Lagrangian for the $a_1\rho\pi$ system must contain at least two terms
of those we have studied. This hampers the factorization of
the cross section and makes the use of formulas (\ref{myformula})
and (\ref{mydecay}) impossible.

\section{Comments and conclusions}
\label{comments}

We have shown that the usual phenomenological way of describing the
invariant mass spectrum of the resonance decay products is not fully 
correct in two respects. 

Firstly, the field theory derivation leads to a factorizing formula for
the cross section that is a little different from that used so far. The 
difference may not be very important from the pragmatic point of view because 
in the close vicinity of the resonant mass the corrections are small.
For masses far from the resonant mass both formulas lose validity anyhow
because so do the expressions for resonance propagators [Eqs.(\ref{scalarprop}) 
and (\ref{vectorprop})]. The validity of our formula has been confirmed 
independently by applying it to a class of electromagnetic processes 
in which a pair of oppositely charged leptons (or quarks) is produced.

The second aspect is more important. We have argued that in some cases
one cannot use a factorizing formula at all. A sufficient condition for
the factorization taking place was shown (\ref{gi}) for vector and
axial-vector resonances. This condition is not satisfied, {\it e.g.},
for $K^*\ra K+\pi$ and $a_1\ra \rho+\pi$ decays.

In this connection a more general question arises how to define the mass
and width of a resonance in the situation when a factorizing
formula cannot be used. The usual procedure of choosing the resonance 
production cross
section as a few-parameter smooth formula and fitting the mass and width 
entering the resonance decay part of (\ref{myformula}) is not applicable any 
longer. We must use a more complete model for the matrix element of the whole 
process spanning from the initial state to the final state containing stable
hadrons which are directly observed in the experiment. The mass and width 
of a resonance thus enter the final mass spectrum formula in a way that is
both different from Eq. (\ref{myformula}) and model dependent. 
In simple cases, when partial wave amplitudes can be determined, one may 
certainly turn for help to analytic methods and define the resonance 
parameters by means of the position of a pole in the complex $s=M^2$ plane. 
But this can be done very rarely due to the complexity of final states in
contemporary high energy experiments.   

In addition to these two problems, which we addressed in this paper in some 
detail, there is  another one, more general, but also ignored very often.
It is the problem of the interference of diagrams when a particle is produced
both in the resonance decay and by other mechanisms. Its
effect may be disastrous and it would be very useful to study the
justification of the resonance formula in such an environment at least
in some simple examples. One study of this kind was done a long time ago
\cite{bergia60} and showed that the interference modified radically the mass 
spectrum in isobaric nucleon model.

All this puts forward a question whether it has sense to define the mass
and width of some resonances ($a_1$, for example) as unique parameters,
independently of the process the resonance takes part in. We suspect
that the parameters of the resonance which does not satisfy the
factorizing condition have a well-defined meaning only if accompanied by 
the specification of the production process in which they were measured
and by the formula which was used to fit the experimental data.

\begin{flushleft}
{\bf Acknowledgement}
\end{flushleft}
This work was completed during the author's visit to CERN, which was
supported by the CERN Theory Division and by the Czech Committee for 
Collaboration with CERN.

\newpage

\begin{figure}[t]
\begin{center}
\leavevmode
\setlength \epsfxsize{5cm}
\epsffile{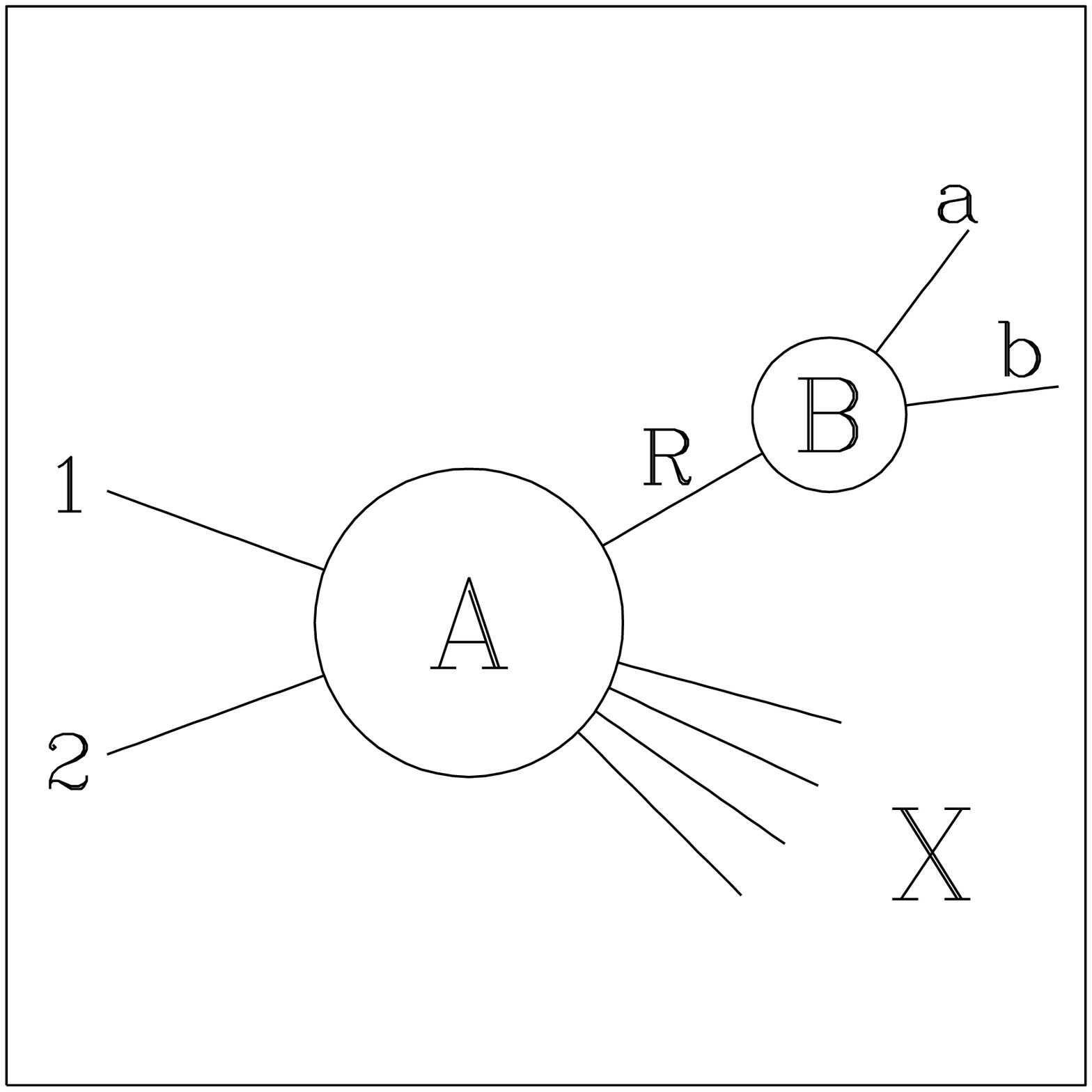}
\end{center}
\caption{Feynman diagram of the reaction \protect{$1+2\ra a+b+X$} under t
he assumptions stated in Sec.~2.1. Symbols $A$ and $B$ are used in Sec.~2.3.}
\end{figure}

\begin{figure}[t]
\begin{center}
\leavevmode
\setlength \epsfxsize{5cm}
\epsffile{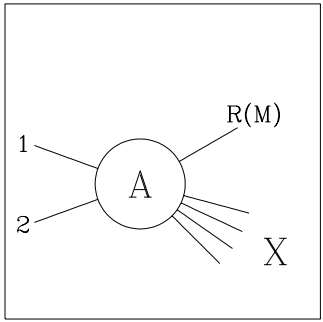}
\end{center}
\caption{Feynman diagram of the reaction \protect{$1+2\ra R(M)+X$}. 
Symbol $A$ is used in Sec.~2.3.}
\end{figure}

\begin{figure}[t]
\begin{center}
\leavevmode
\setlength \epsfxsize{5cm}
\epsffile{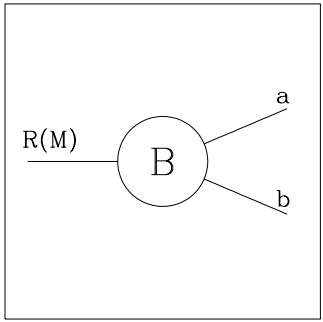}
\end{center}
\caption{Feynman diagram of the decay \protect{$1\ra a+b$}. Symbol  
$B$ is used in Sec.~2.3.}
\end{figure}

\begin{figure}[t]
\begin{center}
\leavevmode
\setlength \epsfxsize{5cm}
\epsffile{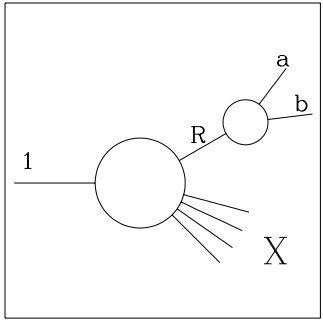}
\end{center}
\caption{Feynman diagram of the decay \protect{$1\ra a+b+X$}.}
\end{figure}

\begin{figure}[t]
\begin{center}
\leavevmode
\setlength \epsfxsize{5cm}
\epsffile{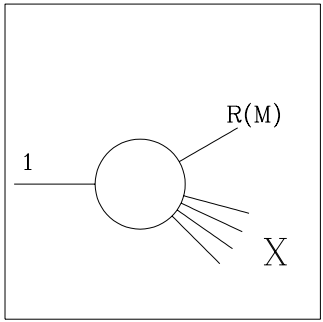}
\end{center}
\caption{Feynman diagram of the decay \protect{$1\ra R(M)+X$}.}
\end{figure}

\begin{figure}[t]
\begin{center}
\leavevmode
\setlength \epsfxsize{5cm}
\epsffile{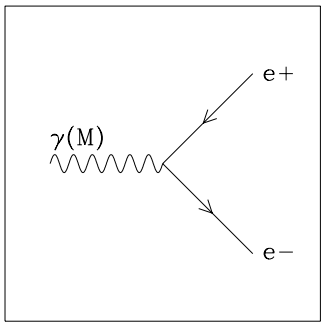}
\end{center}
\caption{Feynman diagram of the ``massive" photon decay
to an $e^+e^-$ pair.} 
\end{figure}

\end{document}